\documentclass[conference]{IEEEtran}

\usepackage{graphicx,amsmath,url}
\usepackage{microtype,lmodern}
\usepackage{mathtools,amssymb,gensymb}
\usepackage{tgtermes} 
\usepackage[T1]{fontenc}
\usepackage[utf8]{inputenc}
\usepackage{color,soul}
\usepackage[colorinlistoftodos]{todonotes}
\usepackage{acronym}
\usepackage{array}
\usepackage{booktabs}
\usepackage[compress]{cite}

\usepackage[normalem]{ulem}
\usepackage[switch]{lineno}
\usepackage{listings}
 
\usepackage{longtable}
\usepackage[bookmarks={false},colorlinks,breaklinks]{hyperref} 
\hypersetup{hypertexnames=true, linkcolor=blue,anchorcolor=black,citecolor=blue,urlcolor=blue}
\graphicspath{{imglocal/}{img/}}
\usepackage[caption=false,font=footnotesize]{subfig}

\acrodef{DR-RZ}[DR-RZ]{Dual-Rail Return-to-Zero}
\acrodef{LEDR}[LEDR]{Level-Encoded Dual-Rail}
\acrodef{PE}[PE]{Phase Encoding}
\acrodef{CML}[CML]{Current Mode Logic}
\acrodef{LVDS}[LVDS]{Low Voltage Differential Signaling}
\acrodef{CDR}[CDR]{Clock-Data Recovery}
\acrodef{ADC}[ADC]{Analog to Digital Converter}
\acrodef{ADEX}[AdExp-I\&F]{Adaptive-Exponential Integrate and Fire}
\acrodef{AER}[AER]{Address-Event Representation}
\acrodef{AEX}[AEX]{AER EXtension board}
\acrodef{AE}[AE]{Address-Event}
\acrodef{AFM}[AFM]{Atomic Force Microscope}
\acrodef{AMDA}[AMDA]{AER Motherboard with D/A converters}
\acrodef{ANN}[ANN]{Attractor Neural Network}
\acrodef{API}[API]{Application Programming Interface}
\acrodef{ARM}[ARM]{Advanced RISC Machine}
\acrodef{ASIC}[ASIC]{Application Specific Integrated Circuit}
\acrodef{BCM}[BMC]{Bienenstock-Cooper-Munro}
\acrodef{BD}[BD]{Bundled Data}
\acrodef{BEOL}[BEOL]{Back-end of Line}
\acrodef{BG}[BG]{Bias Generator}
\acrodef{BMI}[BMI]{Brain-Machince Interface}
\acrodef{CAD}[CAD]{Computer Aided Design}
\acrodef{CAM}[CAM]{Content Addressable Memory}
\acrodef{CAVIAR}[CAVIAR]{Convolution AER Vision Architecture for Real-Time}
\acrodef{CCN}[CCN]{Cooperative and Competitive Network}
\acrodef{CMOL}[CMOL]{``Hybrid CMOS nanoelectronic circuits''}
\acrodef{CMIM}[CMIM]{Metal-insulator-metal Capacitor}
\acrodef{CMOS}[CMOS]{Complementary Metal-Oxide-Semiconductor}
\acrodef{COTS}[COTS]{Commercial Off-The-Shelf}
\acrodef{CPG}[CPG]{Central Pattern Generator}
\acrodef{CPLD}[CPLD]{Complex Programmable Logic Device}
\acrodef{CPU}[CPU]{Central Processing Unit}
\acrodef{CV}[CV]{Coefficient of Variation}
\acrodef{DAC}[DAC]{Digital to Analog Converter}
\acrodef{DAS}[DAS]{Dynamic Auditory Sensor}
\acrodef{DAVIS}[DAVIS]{Dynamic and Active Pixel Vision Sensor}
\acrodef{DBN}[DBN]{Deep Belief Network}
\acrodef{DFA}[DFA]{Deterministic Finite Automaton}
\acrodef{DMA}[DMA]{Direct Memory Access}
\acrodef{DNF}[DNF]{Dynamic Neural Field}
\acrodef{DNN}[DNN]{Deep Neural Network}
\acrodef{DOF}[DOF]{Degrees of Freedom}
\acrodef{DPE}[DPE]{Dynamic Parameter Estimation}
\acrodef{DPI}[DPI]{Differential Pair Integrator}
\acrodef{DRAM}[DRAM]{Dynamic Random Access Memory}
\acrodef{DR}[DR]{Dual Rail}
\acrodef{DSP}[DSP]{Digital Signal Processor}
\acrodef{DVS}[DVS]{Dynamic Vision Sensor}
\acrodef{EBL}[EBL]{Electron Beam Lithography}
\acrodef{EDVAC}[EDVAC]{Electronic Discrete Variable Automatic Computer}
\acrodef{EIN}[EIN]{Excitatory-Inhibitory Network}
\acrodef{EM}[EM]{Expectation Maximization}
\acrodef{EPSC}[EPSC]{Excitatory Post-Synaptic Current}
\acrodef{EPSP}[EPSP]{Excitatory Post-Synaptic Potential}
\acrodef{FD-SOI}[FD-SOI]{Fully-Depleted Silicon on Insulator}
\acrodef{FET}[FET]{Field-Effect Transistor}
\acrodef{FFT}[FFT]{Fast Fourier Transform}
\acrodef{FI}[F-I]{Frequency-Current}
\acrodef{FPGA}[FPGA]{Field Programmable Gate Array}
\acrodef{FSA}[FSA]{Finite State Automaton}
\acrodef{FSM}[FSM]{Finite State Machine}
\acrodef{GOPS}[GOPS]{Giga-Operations per Second}
\acrodef{GPU}[GPU]{Graphical Processing Unit}
\acrodef{GUI}[GUI]{Graphical User Interface}
\acrodef{HAL}[HAL]{Hardware Abstraction Layer}
\acrodef{HH}[H\&H]{Hodgkin \& Huxley}
\acrodef{HMM}[HMM]{Hidden Markov Model}
\acrodef{HRS}[HRS]{High-Resistive State}
\acrodef{HR}[HR]{Human Readable}
\acrodef{HW}[HW]{Hardware}
\acrodef{ICT}[ICT]{Information and Communication Technology}
\acrodef{IC}[IC]{Integrated Circuit}
\acrodef{IF2DWTA}[IF2DWTA]{Integrate \& Fire 2--Dimensional WTA}
\acrodef{IFSLWTA}[IFSLWTA]{Integrate \& Fire Stop Learning WTA}
\acrodef{IF}[I\&F]{Integrate-and-Fire}
\acrodef{IMU}[IMU]{Inertial Measurement Unit}
\acrodef{INCF}[INCF]{International Neuroinformatics Coordinating Facility}
\acrodef{INI}[INI]{Institute of Neuroinformatics}
\acrodef{IO}[I/O]{Input/Output}
\acrodef{IPSC}[IPSC]{Inhibitory Post-Synaptic Current}
\acrodef{IPSP}[IPSP]{Inhibitory Post-Synaptic Potential}
\acrodef{IP}[IP]{Intellectual Property}
\acrodef{ISI}[ISI]{Inter-Spike Interval}
\acrodef{JFLAP}[JFLAP]{Java - Formal Languages and Automata Package}
\acrodef{LFP}[LFP]{Local Field Potential}
\acrodef{LNA}[LNA]{Low-Noise Amplifier}
\acrodef{LPF}[LPF]{Low-Pass Filter}
\acrodef{LRS}[LRS]{Low-Resistive State}
\acrodef{LSM}[LSM]{Liquid State Machine}
\acrodef{LTD}[LTD]{Long Term Depression}
\acrodef{LTI}[LTI]{Linear Time-Invariant}
\acrodef{LTP}[LTP]{Long Term Potentiation}
\acrodef{LTU}[LTU]{Linear Threshold Unit}
\acrodef{LUT}[LUT]{Look-Up Table}
\acrodef{MCMC}[MCMC]{Markov-Chain Monte Carlo}
\acrodef{MEMS}[MEMS]{Micro Electro Mechanical System}
\acrodef{MIM}[MIM]{Metal Insulator Metal}
\acrodef{MOS}[MOS]{Metal Oxide Semiconductor}
\acrodef{MOSCAP}[MOSCAP]{Metal Oxide Semiconductor Capacitor}
\acrodef{MOSFET}[MOSFET]{Metal Oxide Semiconductor Field-Effect Transistor}
\acrodef{MRI}[MRI]{Magnetic Resonance Imaging}
\acrodef{NDFSM}[NDFSM]{Non-deterministic Finite State Machine} 
\acrodef{ND}[ND]{Noise-Driven}
\acrodef{NEF}[NEF]{Neural Engineering Framework}
\acrodef{NHML}[NHML]{Neuromorphic Hardware Mark-up Language}
\acrodef{NIL}[NIL]{Nano-Imprint Lithography}
\acrodef{NMDA}[NMDA]{N-Methyl-D-Aspartate}
\acrodef{NME}[NE]{Neuromorphic Engineering}
\acrodef{OTA}[OTA]{Operational Transconductance Amplifier}
\acrodef{PCB}[PCB]{Printed Circuit Board}
\acrodef{PCHB}[PCHB]{Pre-Charge Half-Buffer}
\acrodef{PSC}[PSC]{Post-Synaptic Current}
\acrodef{PFM}[PFM]{Pulse Frequency Modulation}
\acrodef{PSTH}[PSTH]{Peri-Stimulus Time Histogram}
\acrodef{QDI}[QDI]{Quasi-Delay Insensitive}
\acrodef{RAM}[RAM]{Random Access Memory}
\acrodefplural{RAM}[RAMs]{Random Access Memories}
\acrodef{RMSE}[RMSE]{Root Mean Squared-Error}
\acrodef{RMS}[RMS]{Root Mean Squared}
\acrodef{RNN}[RNN]{Recurrent Neural Network}
\acrodef{ROLLS}[ROLLS]{Reconfigurable On-Line Learning Spiking}
\acrodef{RRAM}[RRAM]{Resistive Random Access Memory}
\acrodefplural{RRAM}[R-RAMs]{Resistive Random Access Memories}
\acrodef{SAC}[SAC]{Selective Attention Chip}
\acrodef{SCX}[SCX]{Silicon CorteX}
\acrodef{SD}[SD]{Signal-Driven}
\acrodef{SEM}[SEM]{Spike-based Expectation Maximization}
\acrodef{SLAM}[SLAM]{Simultaneous Localization and Mapping}
\acrodef{SOI}[SOI]{Silicon on Insulator}
\acrodef{SOC}[SOC]{System-On-Chip}
\acrodef{SRAM}[SRAM]{Static Random Access Memory}
\acrodef{STDP}[STDP]{Spike-Timing Dependent Plasticity}
\acrodef{STD}[STD]{Short-Term Depression}
\acrodef{STP}[STP]{Short-Term Plasticity}
\acrodef{STT}[STT]{Spin-Transfer Torque}
\acrodef{STT-MRAM}[STT-MRAM]{Spin-Transfer Torque Magnetic Random Access Memory}
\acrodefplural{STT-MRAM}[STT-MRAMs]{Spin-Transfer Torque Magnetic Random Access Memories}
\acrodef{SW}[SW]{Software}
\acrodef{TFT}[TFT]{Thin Film Transistor}
\acrodef{USB}[USB]{Universal Serial Bus}
\acrodef{VHDL}[VHDL]{VHSIC Hardware Description Language}
\acrodef{VLSI}[VLSI]{Very Large Scale Integration}
\acrodef{VOR}[VOR]{Vestibulo-Ocular Reflex}
\acrodef{WTA}[WTA]{Winner-Take-All}
\acrodef{XML}[XML]{eXtensible Mark-up Language}
\acrodef{divmod3}[DIVMOD3]{divisibility of a number by 3}
\acrodef{hWTA}[hWTA]{Hard Winner-Take-All}
\acrodef{sWTA}[sWTA]{soft Winner-Take-All}

\begin{document}
%
\title{A bi-directional Address-Event transceiver block for low-latency inter-chip communication in neuromorphic systems}

\author{\IEEEauthorblockN{Ning Qiao and Giacomo~Indiveri}
\IEEEauthorblockA{Institute of Neuroinformatics, University of Zurich and ETH Zurich, 
Zurich, Switzerland\\
Email: [qiaoning$|$giacomo]@ini.uzh.ch}}


\maketitle

\begin{abstract}
  Neuromorphic systems typically use the \ac{AER} to transmit signals among nodes, cores, and chips. Communication of \acp{AE} between neuromorphic cores/chips typically requires two parallel digital signal buses for \ac{IO} operations. This requirement can become very expensive for large-scale systems in terms of both dedicated \ac{IO} pins and power consumption.
  In this paper we present a compact fully asynchronous event-driven transmitter/receiver block that is both power efficient and \ac{IO} efficient. This block implements high-throughput low-latency bi-directional communication through a parallel \ac{AER} bus. We show that by placing the proposed \ac{AE} transceiver block in two separate chips and linking them by a single \ac{AER} bus, we can drive the communication and switch the transmission direction of the shared bus on a single event basis, from either side with low-latency. We present experimental results that validate the circuits proposed and demonstrate reliable bi-directional event transmission with high-throughput. The proposed \ac{AE} block, integrated in a neuromorphic chip fabricated using a 28\,nm FDSOI process, occupies a silicon die area of 140\,$\mu$m~$\times$~70\,$\mu$m. The experimental measurements show that the event-driven \ac{AE} block combined with standard digital \acp{IO} has a direction switch latency of 5\,ns and can achieve a worst-case bi-directional event transmission throughput of 28.6\,M$\cdot$Events/second while consuming 11\,pJ per event (26-bit) delivery.  
\end{abstract}

\IEEEpeerreviewmaketitle

\acresetall

\section{Introduction}
\label{sec:introduction}

The \ac{AER} has been widely used in brain-inspired neuromorphic systems as a communication protocol for transmitting and receiving spikes encoded as \acp{AE} among spiking silicon neurons and synapses. For example dynamic vision sensors~\cite{Delbruck_etal10a} and silicon cochleas~\cite{Liu_etal14a} use the \ac{AER} to transmit their sensory processing outputs to \acp{AER} neuromorphic processors and transceivers~\cite{Park_etal14,Merolla_etal14a,Furber_etal14,Qiao_etal15,Moradi_etal17}. As these types of neuromorphic VLSI systems typically require \acp{AE} to be transmitted with high throughput and low latency, the strategy employed to implement the communication protocol makes use of asynchronous bit-parallel \ac{AER} channels. This strategy however is not scalable, as the width of the parallel bus and the power required to transmit these parallel events scales with the size of the network. In addition, the pin count and power requirements become even larger if one desires to build modular systems with north/south, east/west \ac{IO} links necessary to tile multiple cores or chips in 2D arrays~\cite{Merolla_etal14a,Moradi_etal17}. Instead of simple pure parallel \ac{AER} protocol, some approaches use a ``word-serial'' protocol to transmit multiple row addresses for every column address serviced (or vice-versa) to reduce pin numbers~\cite{Boahen04,Brandli_etal14}. Furthermore, bit-serial \ac{LVDS} \ac{AER} has been proposed as a potential solution to transmit events in a fully bit-serial format to further reduce pin numbers~\cite{Zamarreno-Ramos_etal13a}. However, these approaches lead to significant increment in latency and overhead for the complexity of the circuit implementation. Moreover, the design proposed in~\cite{Zamarreno-Ramos_etal13a} needs additional clock generation and synchronization circuits which is expensive for fully asynchronous neuromorphic system. 

In this paper, we present a compact fully-asynchronous event-driven \ac{AE} transceiver block which can be easily combined with standard digital \acp{IO} to realize bi-directional inter-chip \ac{AE} communication through a single parallel \ac{AER} bus with high-throughput and low-latency. In the next Section, we introduce architecture of the proposed \ac{AE} transceiver block. In Section~\ref{sec:circuits}, we describe the circuits that implement the proposed bi-directional \ac{AER} block. In Section~\ref{sec:experimental-results} we present experimental results obtained from the measurements of a test chip fabricated in 28\,nm FDSOI process. We present concluding remarks and discussion in Section~\ref{sec:conclusions}.

\section{Architecture}
\label{sec:architecture}

\begin{figure}
  \centering
  \includegraphics[width=0.47\textwidth]{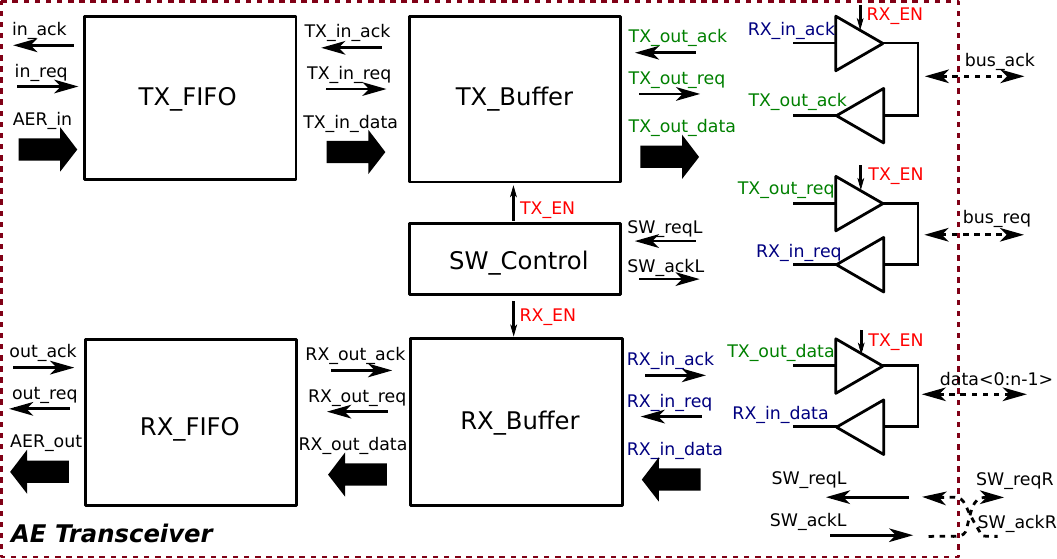}
  \caption{Architecture of proposed bi-directional \ac{AE} transceiver block. SW\_Control block checks states of two linked chip and generates control signal $TX/RX\_EN$ to allow TX\_Buffer to push events on signal AER bus or allow RX\_Buffer to take events from the single AER bus. Bi-directional tri-state buffers are switched by $TX/RX\_EN$ for bus direction. TX/RX\_FIFOs are added to increase throughput of proposed AE transmission block.}
    \label{fig:arch}
\end{figure}

\begin{table}
  \begin{center}
  \begin{tabular}{|c | c | c | c|}
    \hline
    SW\_reqL (SW\_ackR) & SW\_ackL (SW\_reqR)  & Left Mode  & Right Mode  \\
    \hline
    0 & 1 & TX & RX\\
    $\Uparrow$ & 1 & TX & RX\\
    1 & $\Downarrow$ & TX$\to$RX & RX$\to$TX\\
    1 & $\Uparrow$ & RX & TX\\
    $\Downarrow$ & 1 & RX$\to$TX & TX$\to$RX  \\
    \hline 
  \end{tabular} 
  \end{center}
  \caption{SW\_req/ack states for mode switching }
  \label{table:code} 
\end{table}

\begin{figure}
  \centering
  \includegraphics[width=0.48\textwidth]{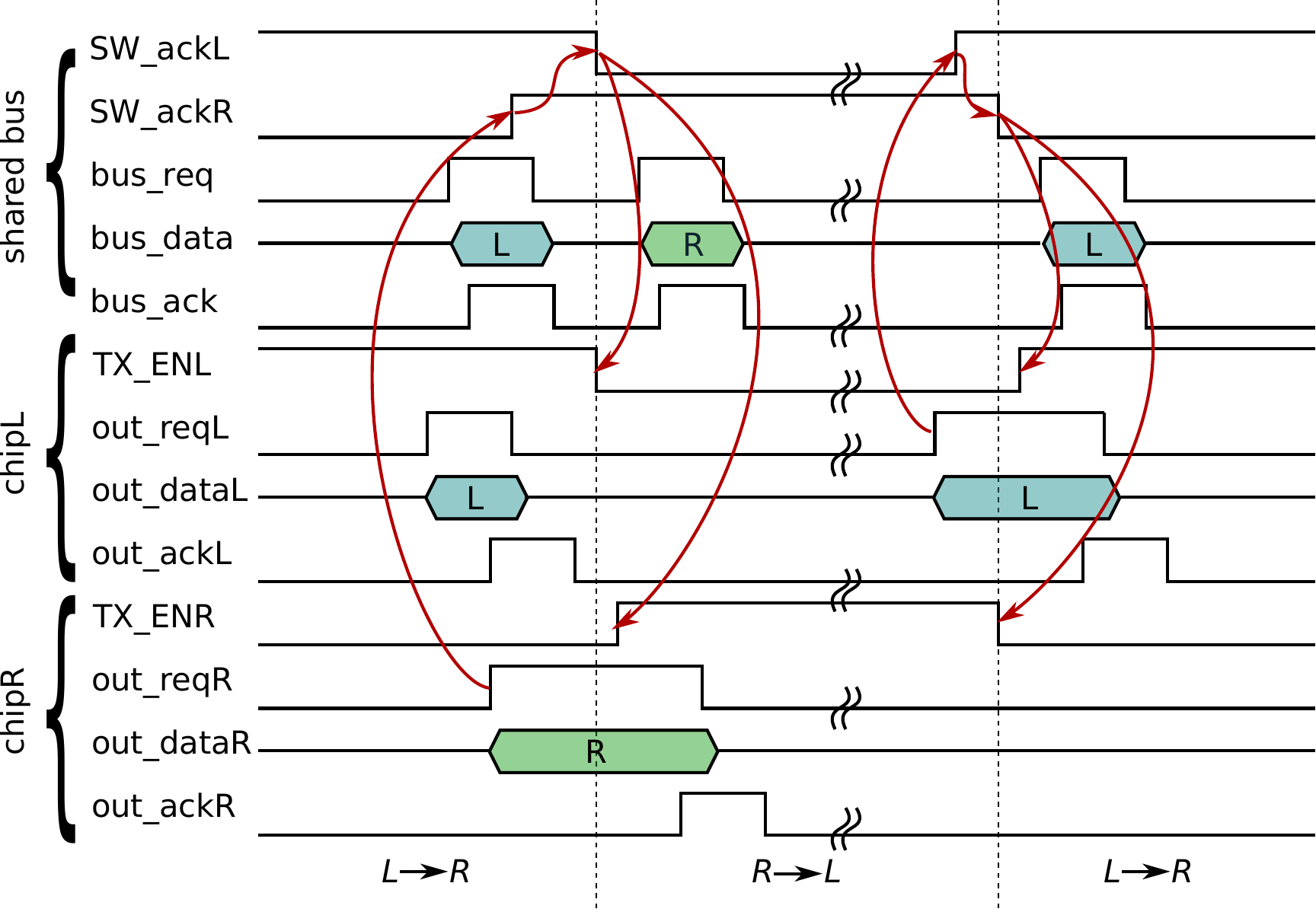}
  \caption{Mode switch scheme of proposed AER in/out block for bi-directional events transmission.}
  \label{fig:scheme}
\end{figure}

Figure~\ref{fig:arch} shows the architecture of the proposed \ac{AE} transceiver block. Bi-directional chip communication can be implemented by connecting two \ac{AE} transceiver blocks directly with a single shared bit-parallel AER bus. As is shown in Fig.~\ref{fig:arch}, $SW\_ack$ and $SW\_req$ from two linked AE blocks are swapped and connected to announce state of each other. The SW\_Control block in each \ac{AE} transceiver block then checks the states of two connected chip which are indicated by $SW\_ack$ and $SW\_req$, and generates control signals $TX/RX\_EN$ to allow TX\_Buffer to push events on the shared AER bus or allow RX\_Buffer to take events from the bus. Each AE block use $SW\_ack$ to identify its own states (i.e., logic ``1'' if this AE block need to switch to transmitter mode ``TX'' for transmitting events, and logic ``0'' if currently this AE block has no event to transmit and can be switched to receiver mode ``RX''), and use $SW\_req$ to get states of it's linked AE block. SW\_Control block on both sides will generate control signals $TX\_EN$ and $RX\_EN$ to switch on/off TX\_Buffer and RX\_Buffer and alternatively map terminals of TX\_Buffer or RX\_Buffer to the shared bus for a mode switching. Table~\ref{table:code} shows how modes are switched in different cases, with $\Uparrow$ representing logic ``0'' to ``1'' and $\Downarrow$ representing logic ``1'' to ``0''. 

Moreover, conditions need to be met for a safe mode switching. An AE transceiver block will only request a mode switching RX$\to$TX by asserting its SW\_ack $\Uparrow$ when: 1) The block is currently in a ``RX'' mode; 2)The block has received at least one event in ``RX'' mode (except that this block is initially reset to ``RX'' mode for a chip-level global reset);  And 3) one or more events need to be transmitted.  An \ac{AE} transceiver block will only acknowledge a mode switching request from its lined \ac{AE} block by de-asserting its SW\_ack $\Downarrow$ when: 1) The block is currently in a ``TX'' mode, and 2) it received a mode switching request.

Figure~\ref{fig:scheme} shows an example how bi-directional transmission is implemented with proposed AE transceiver block following 4-phase handshaking. Assume that two AE transceiver blocks are linked by a signal AER bus, and initially we set $SW\_ackL$ of left block to logic ``1'' and $SW\_ackR$ of right block to logic ``0''. So that initially the left AE block is in ``TX'' mode and the right AE block is in ``RX'' mode, to allow event transmission from left to right. Once there is an event need to be transmitted on right side, $SW\_ackR$ will be assert to ``1'' to request a mode switching.  After requested by $SW\_ackR$, as soon as there is no more event need to be transmitted by the left AE block, $SW\_ackL$ will be deassert to ``0'' to acknowledge the mode switching request. Correspondingly, $TX/RX\_EN$ in both blocks will be flipped to complete the mode switching. 

Bi-directional tri-state buffers as shown in Fig.~\ref{fig:arch} are then switched by $TX/RX\_EN$ for mapping signals of TX/RX\_Buffer blocks to shared AER bus. We should notice that the tri-state buffers can be directly replaced with standard digital \ac{IO} with $TX/RX\_EN$ as a configure signal for Input/Output switching. Input and output FIFOs are added to increase throughput of proposed AE transceiver block.

\section{Circuits Implementation}
\label{sec:circuits}

\begin{figure}
  \centering
  \includegraphics[width=0.48\textwidth]{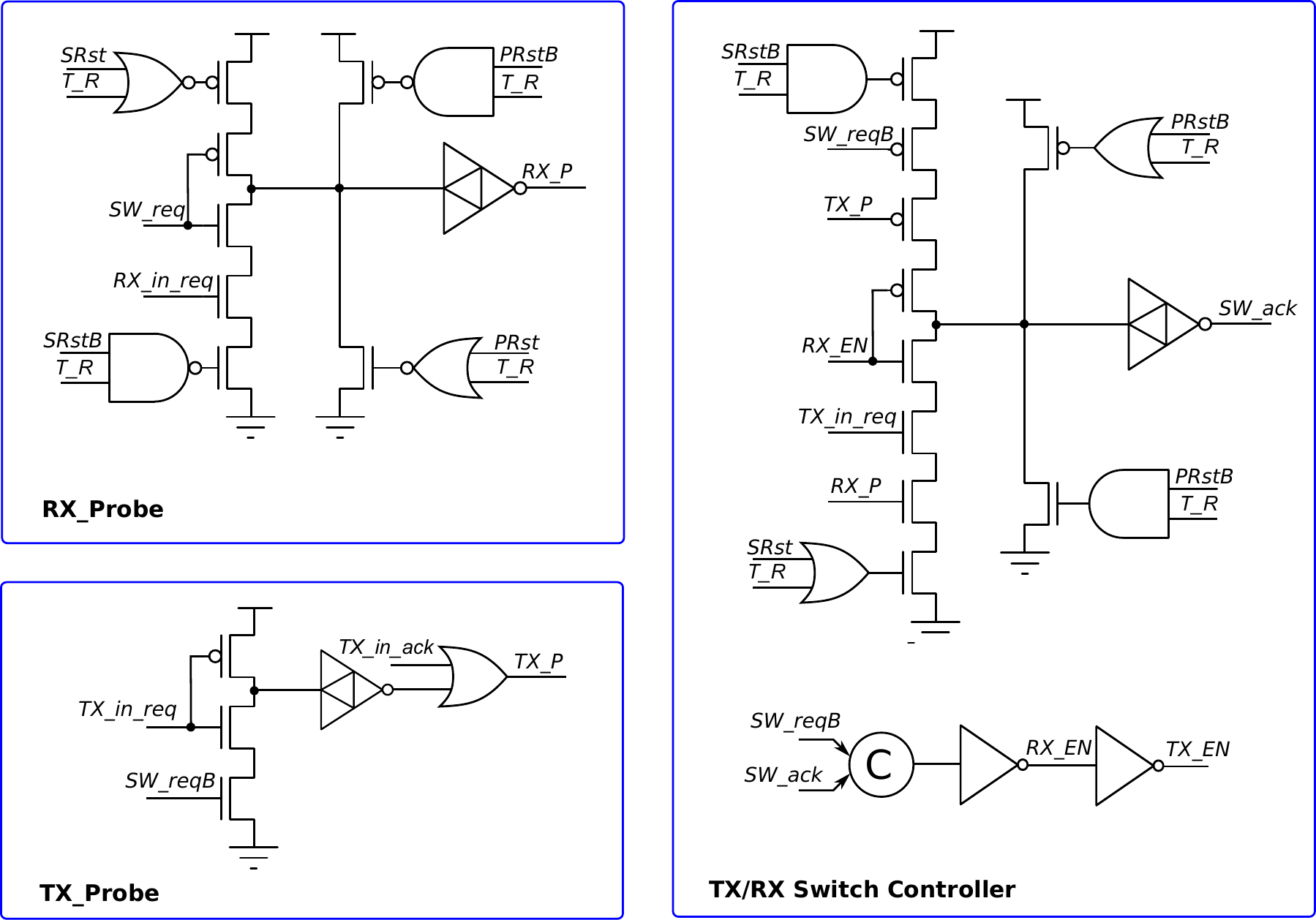}
  \caption{Circuit implementation of SW\_Control block. RX\_Probe and TX\_Probe are used to probe its state once it is in ``RX'' mode or ``TX'' mode. TX/RX Switch Control generates SW\_ack signal and further generate TX/RX\_EN for its mode switch.}
  \label{fig:SW_Control}
\end{figure}

The proposed \ac{AE} transceiver block is implemented following 4-phase handshaking protocol based on \ac{PCHB}. Figure~\ref{fig:SW_Control} shows circuit implementation of SW\_Control block for controlling mode switching. RX\_Probe is employed to probe whether the belonging AE block has received at least one event as a receiver ($PX\_P$ = ``1'') in ``RX'' mode ($SW\_req$ = ``1'').  TX\_Probe is used to probe whether currently the belonging AE transceiver block has no event to be transmitted ($TX\_P$ = ``0'') as a transmitter in ``TX'' mode ($TX\_EN$ = ``1'') when its linked AE transceiver block requests to switch the mode ($SW\_req \Uparrow$). Switch Controller sub-block requests a mode switching RX$\to$TX (by asserting $SW\_ack$ to ``1'') when  a coming event needs to be transmitted ($TX\_in\_req$ = ``1'') if its belonging AE transceiver block is currently in ``RX'' mode ($RX\_EN$ = ``1'') and it has successfully received at least one event ($PX\_P$ = ``1'') in ``RX'' mode. Three NFETs in Switch Controller sub-block gated by $TX\_in\_req$,  $RX\_EN$ and $PX\_P$ implement these guards. Switch Controller block also acknowledges a mode switching request from its linked AE transceiver block for a mode switching TX$\to$RX ($SW\_req$ = ``1'') if currently no event needs to be transmitted ($TX\_P$ = ``0''). Two p-FETs in Switch Controller sub-block gated by $SW\_reqB$ and $TX\_P$ implement these guards. 

As described in previous section, If the AE transceiver block requested a mode switching RX$\to$TX and its linked AE block has acknowledged this request ($SW\_ack$ = ``1'' $\cap$ $SW\_req$ = ``0'') , this AE block will be switched to ``TX'' mode ($TX\_EN$ = ``1''). Otherwise, if this AE transmission block has acknowledged a mode switching TX$\to$RX requested by its linked AE block  ($SW\_req$ = ``1'' $\cap$ $SW\_ack$ = ``0''), this AE block will be switched to ``RX'' mode ($RX\_EN$= ``1''). In these figures, signal ends with ``B'' represent reversed signal. Logic gates gated by $SRst$, $PRst$ are global reset signals used to reset TX and RX Probes to an initial state, for example, $RX\_P$ is reset to ``0'' for ``TX'' mode or ``1'' for ``RX'' mode. 

\begin{figure}
  \centering
  \includegraphics[width=0.48\textwidth]{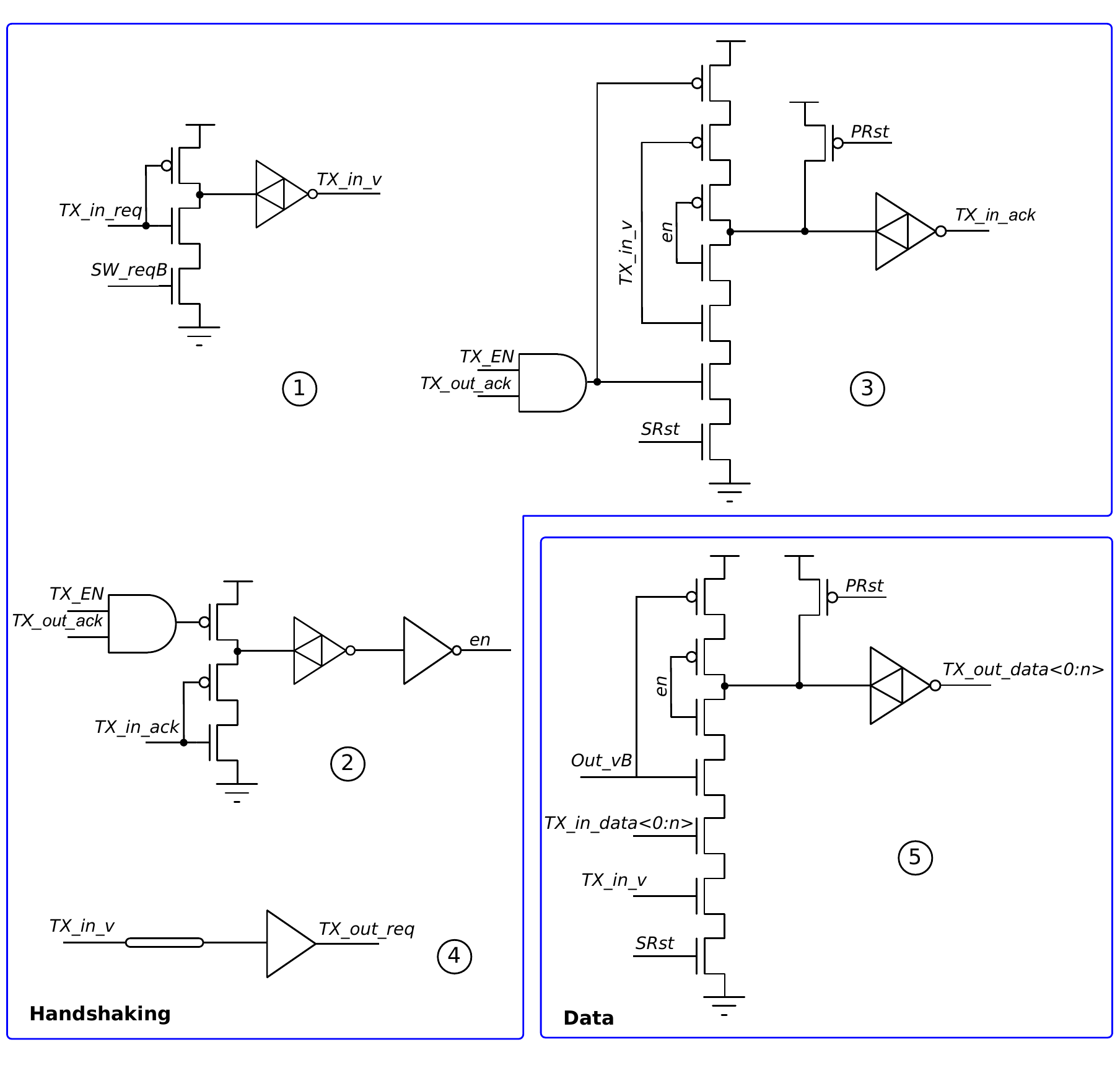}
  \caption{Circuit implementation of 4-phase bundled-data TX\_Buffer based on PCHB. The TX\_Buffer includes handshaking and data blocks. }
  \label{fig:output}
\end{figure}

Figure~\ref{fig:output} shows transistor level circuit implementation of TX\_Buffer based on \ac{PCHB}, following 4-phase bundled-data handshaking protocol. The process stage includes Handshaking and Data function blocks. Block \textcircled{1} guarantees that the process stage only deal with coming events while the linked AE transceiver block is free ($SW\_req$ = ``0''). Block \textcircled{2} checks whether the processing progress is completed to generate handshaking signal for previous process stage. Block \textcircled{3} generates internal enable signal $en$ to enable functional processing.  Matched delay element \textcircled{4} is added to provide the worst case latency of buffer operation from valid input event data to output event.  Block \textcircled{5} implements an event buffer function. 

\begin{figure}
  \centering
  \includegraphics[width=0.48\textwidth]{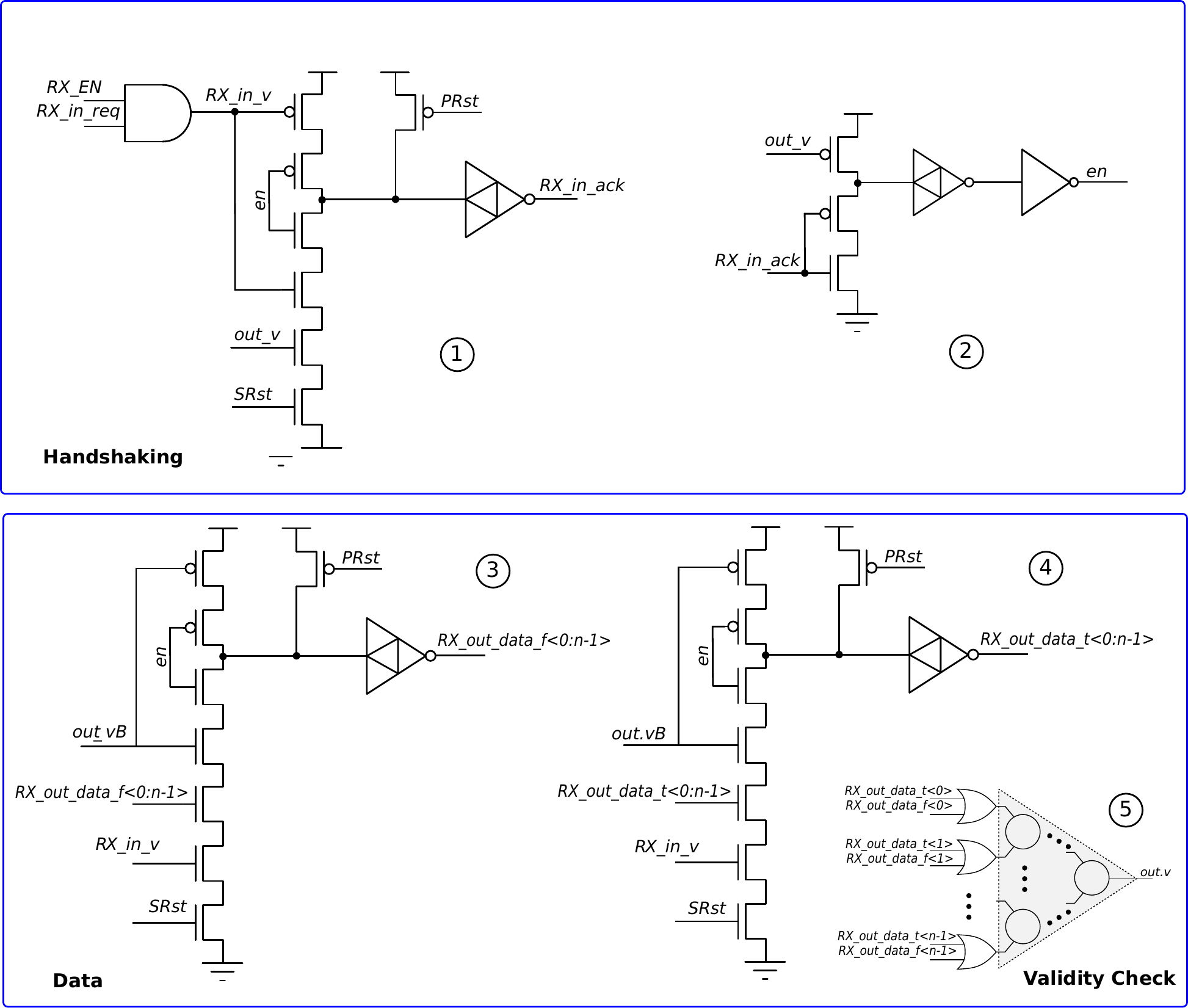}
  \caption{Circuit implementation of 4-phase dual-rail RX\_Buffer based on PCHB. The RX\_Buffer includes handshaking, data and validity check blocks.}
  \label{fig:input}
\end{figure}

RX\_Buffer following 4-phase rail-rail handshaking protocol based on \ac{PCHB} is shown in Fig.~\ref{fig:input}. Block \textcircled{1} checks whether the processing progress is completed and generates acknowledge signal $RX\_in\_ack$ to acknowledge previous process stage for a valid input and completed valid output. Block \textcircled{2} generates internal enable signal $en$ to enable functional processing. Dual-rail protocol (block  \textcircled{3} and \textcircled{4} ) is utilized in this RX\_Buffer and following RX\_FIFO stage for \ac{QDI} processes. Validity check block  \textcircled{5} is employed to indicate output data from this process stage is valid.

\section{Experimental Results}
\label{sec:experimental-results}

\begin{figure}
  \centering
  \includegraphics[width=0.47\textwidth]{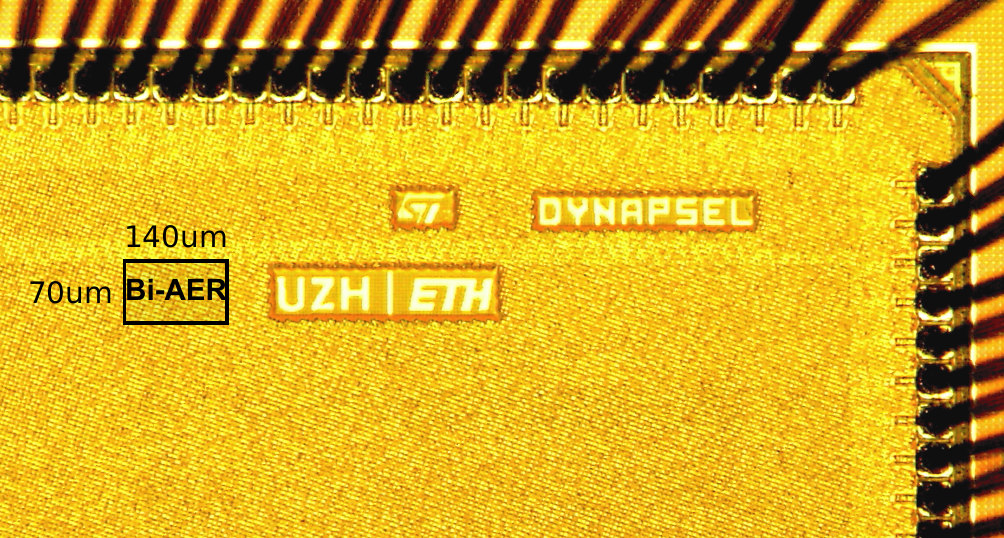}
  \caption{Neuromorphic chip implemented in 28\,nm FDSOI with proposed AE transceiver block combined with standard digital IOs for bi-directional inter-chip AER communication. Each AE block occupies a silicon area of 140\,$\mu$m $\times$ 70\,$\mu$m.}
  \label{fig:die_photo}
\end{figure}

The proposed AE transceiver block is implemented and placed at all chip boarders of a neuromorphic chip in 28\,nm FDSOI process \cite{Qiao_Indiveri16} for implementing 2D chip-array bi-directional 26-bit \ac{AER} communication. Standard digital \ac{IO}s with driven ability of 2\,mA are adopted and internally configured by $TX\_EN$ and $RX\_EN$ for switching event transmission direction. As is shown in Fig.~\ref{fig:die_photo}, each AE block occupies a silicon area of 140\,$\mu$m $\times$ 70\,$\mu$m. By easily utilizing proposed AE blocks, we saved 100 \ac{IO}s which is a significant reducing for a prototype chip with totally 180 \ac{IO}s. 

\begin{figure}
  \centering
  \includegraphics[width=0.45\textwidth]{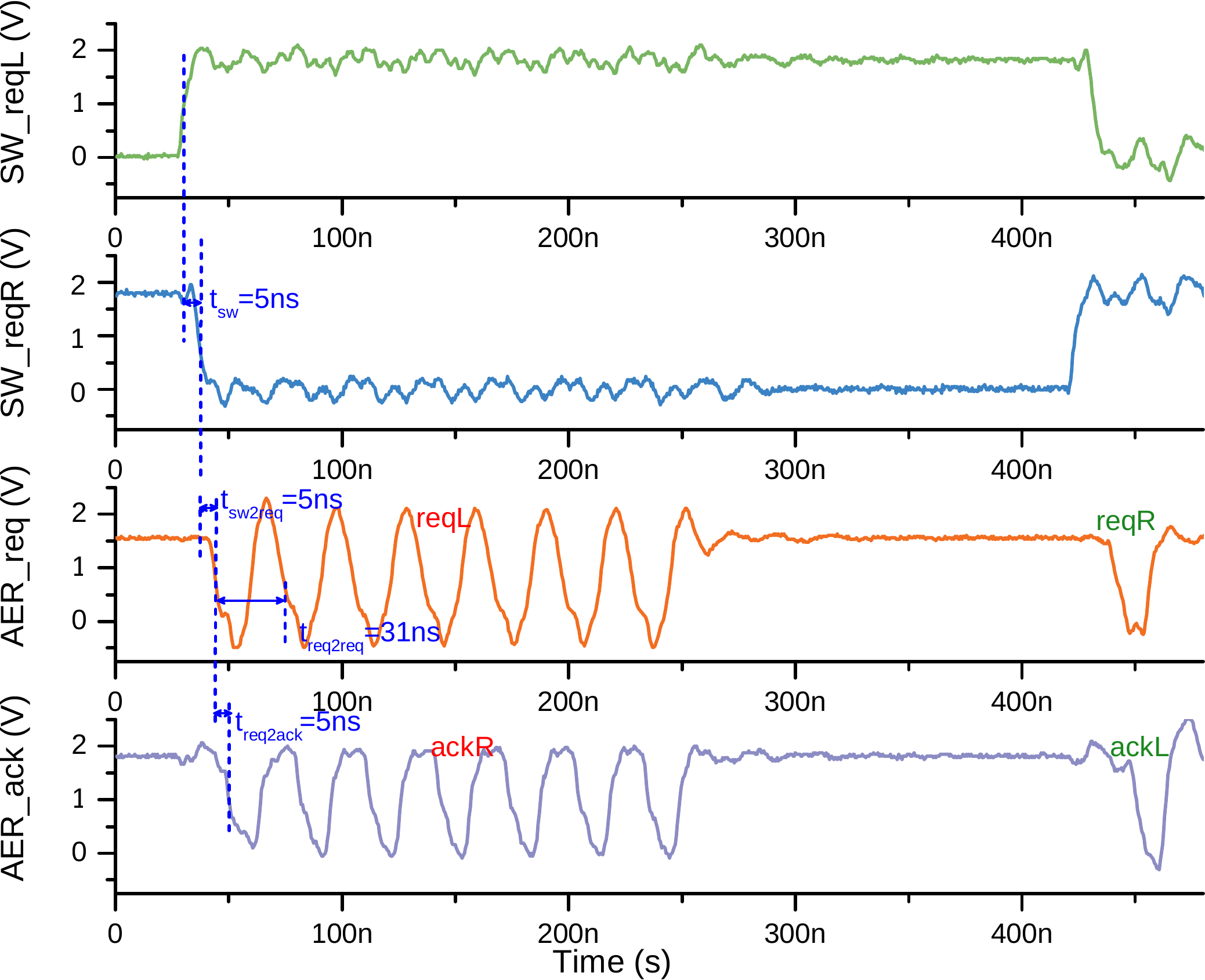}
  \caption{Signal waves for continuously one-direction events transmission with a throughput of 32.3\,M$\cdot$Events/second.}
  \label{fig:experiment2}
\end{figure}

In order to judge the performance, we first measured single direction events transmission performance by continuously sending address events from single direction. As is shown in Fig.~\ref{fig:experiment2}, AE transceivers from two linked chips are first reset to transmission direction from right to left. For continuously events communing from left, AE blocks first need to switch transmission direction with a switching latency $t_{sw}$ of around 5\,ns. Latency from a successfully mode switching to asserting the first request $t_{sw2req}$ is around 5\,ns. For continuously single direction events transmission, latency between two requests $t_{req2req}$ is around 31\,ns, with a throughput of 32.3\,M$\cdot$Events/second. 

\begin{figure}
  \centering
  \includegraphics[width=0.45\textwidth]{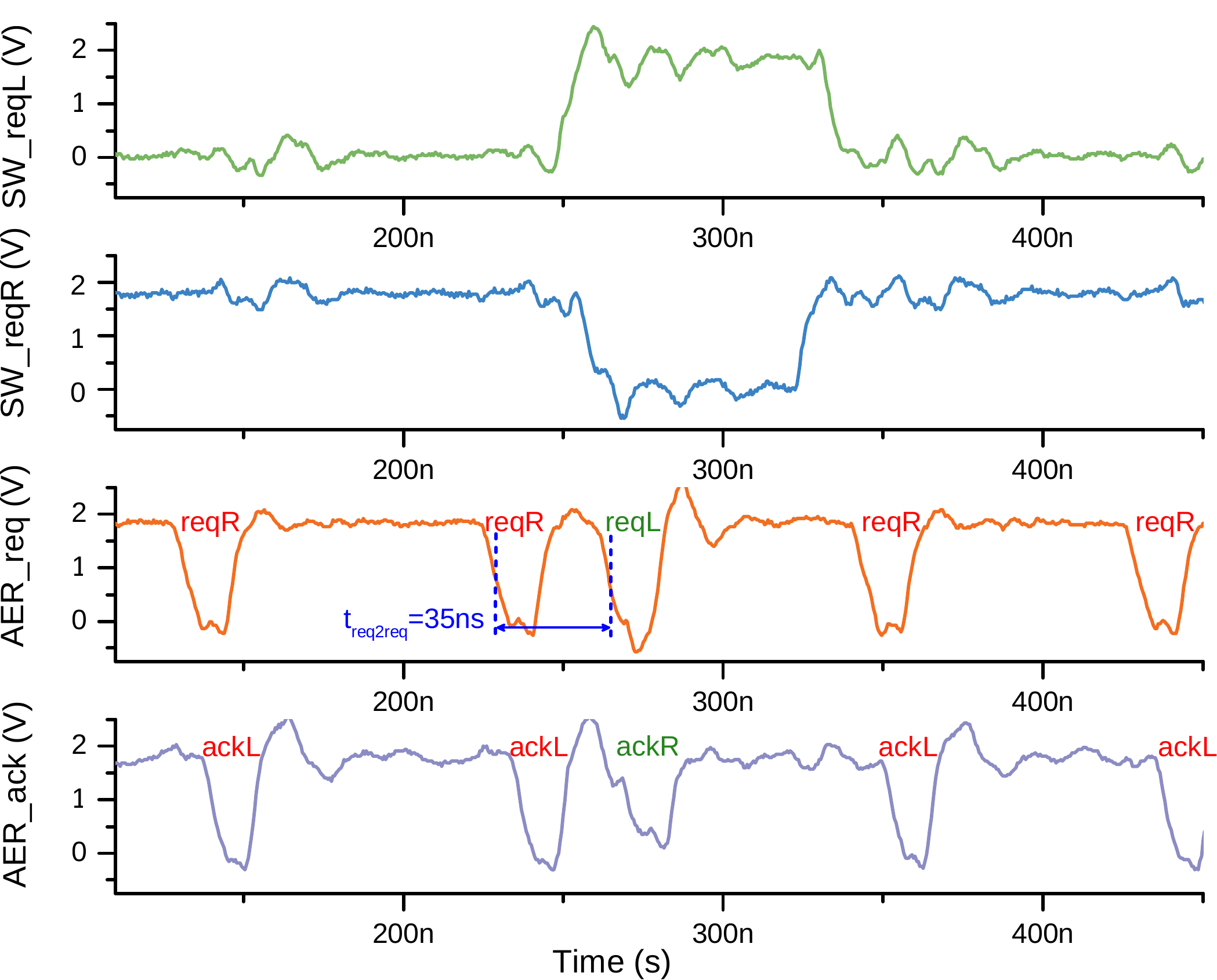}
  \caption{Signal waves for bi-directional events transmission with a throughput of 28.6\,M$\cdot$Events/second.}
  \label{fig:experiment1}
\end{figure}

For bi-directional transmission, we transmitted events from both directions of two linked \ac{AE} blocks. As is shown in Fig.~\ref{fig:experiment1}, request latency of two events from two directions is around 35\,ns, with an achieved worst case bi-directional throughput of 28.6\,M$\cdot$Events/second. Energy for delivering one 26-bit event is 11\,pJ at 1\,V power supply, excluding power consumption from digital IOs.

A summary of the key figures of the proposed AE transceiver block is shown in Table~\ref{table:AE}. 

\begin{table}
  \caption[]{\ac{AE} bi-directional transmission block circuit key figures.}  
  \label{table:AE} 
  \centering
  \begin{tabular}{l  l}  
    \toprule
    Process Technology &  28\,nm FDSOI\\      
    Silicon Area  &  140\,$\mu$m $\times$ 70\,$\mu$m   \\   
    Throughput  (with IO)&  32.3\,MEvents/s / 28.6\,MEvents/s (bi-directional)\\     
    Latency  &  5\,ns  \\     
    Energy per Event (26-bit) &  11\,pJ@1\,V  \\     \midrule       
  \end{tabular} 
\end{table} 

\section{Conclusions}
\label{sec:conclusions}

We presented a compact low-power event-driven bi-directional \ac{AE} transceiver block for high-throughput and low-latency bi-directional inter-chip communication. The proposed fully asynchronous \ac{AE} block is compatible with standard digital \ac{IO}s for easily implementing bi-directional inter-chip communication while saving half \ac{IO}s, comparing with normal bit-parallel \ac{AER} protocol.
Furthermore, it is possible to combine proposed scheme with "sub-words" to further reduce \ac{IO} numbers and power consumption.
We designed and fabricated the proposed \ac{AE} transmission block in 28\,nm FDSOI process with an area of 140\,$\mu$m $\times$ 70\,$\mu$m. Combined with standard digital \ac{IO}s, we implemented 2D spiking neural network bi-directional chip-array communication. Chip measurements show that the proposed \ac{AE} transceiver block can achieve a worst case bi-directional event throughput of 28.6\,M$\cdot$Events/s with energy per event 11\,pJ at 1\,V supply voltage. The latency for switching transmission direction between two AE transmission blocks is around 5\,ns. 

\section*{Acknowledgment}
This work is supported by the EU ERC 
grant ``NeuroP'' (257219) and by the EU ICT grant ``NeuRAM$^3$'' (687299). 

\bibliographystyle{IEEEtran}

\end{document}